\documentstyle[twoside,fleqn,espcrc2,epsf]{article}

\def\zbz{\bar\zeta\zeta}

\def\azbz{\langle\bar\zeta\zeta\rangle}
\def\acbc{\langle\bar\chi\chi\rangle}
\def\lan{\left\langle}

\def\ranf{\right\rangle_{\beta, m}}

\newcommand{\AmS}{{\protect\the\textfont2
  A\kern-.1667em\lower.5ex\hbox{M}\kern-.125emS}}

\title{Dirac Spectrum, Axial Anomaly and the QCD Chiral Phase
Transition
\thanks{This work was done in collaboration with Dong Chen,
Weonjong Lee, Robert Mawhinney, and Decai Zhu. The work was
supported in part by the US department of Energy.} }

\author{Shailesh Chandrasekharan and Norman Christ \\
\vskip 0.1in
Department of Physics, Columbia University, \\
New York, NY 10027, USA}
\begin{document}

\def\thepage{CU--TP--711 \ \ \  hep-lat/9509095}
\thispagestyle{myheadings}

\begin{abstract}

The QCD phase transition is studied on $16^3$ and $32^3 \times 4$
lattices both with and without quark loops.  We introduce a new
zero-flavor or quenched species of quark $\zeta$ and study the
resulting chiral condensate, $\azbz$ as a function of the $\zeta$
mass, $m_\zeta$.  By examining $\azbz$ for $10^{-10} \le m_\zeta
\le 10$ we gain considerable information about the spectrum of
Dirac eigenvalues.  A comparison of $ma=0.01$ and 0.025 shows
little dependence of the Dirac spectrum on such a light,
dynamical quark mass, after an overall shift in $\beta$ is
removed.   The presence of sufficient small eigenvalues to
support anomalous chiral symmetry breaking in the high
temperature phase is examined quantitatively.  In an effort to
enhance these small eigenvalues, $\azbz$ is also examined in the
pure gauge theory in the region of the deconfinement transition
with unexpected results.  Above the critical temperature, the
three $Z_3$ phases show dramatically different chiral behavior.
Surprisingly, the real phase shows chiral symmetry, suggesting
that a system with one flavor of staggered fermion at $N_t=4$
will possess a chiral a phase transition---behavior not expected
in the continuum limit.
\end{abstract}

\maketitle

\section{INTRODUCTION}

The spectrum of the Dirac operator is closely related to a number
of important aspects of finite temperature QCD.  First, a
non-zero density of small eigenvalues is necessary for
spontaneous chiral symmetry breaking\cite{Banks}.  Second, if the
QCD phase transition is second order, one expects this eigenvalue
spectrum to show critical behavior for $T\approx T_c$.  Third,
the Dirac spectrum should be particularly sensitive to the
effects of quark loops since the Dirac determinant for small
quark mass strongly suppresses gauge configurations with small
Dirac eigenvalues.  Finally, the universality arguments which
predict the order of the QCD phase transition\cite{Wilczek}, rely on
the explicit breaking of the continuum $U_A(1)$ symmetry by the
axial anomaly.  Such anomalous symmetry breaking requires a
significant density of small eigenvalues.

Since it is difficult to obtain the information about the
spectrum directly we choose to examine a less familiar quantity:
the quenched or zero-flavor chiral condensate for an auxiliary
quark field  $\zeta(x)$ with mass $m_\zeta$.  By studying the
dependence on $m_\zeta$, we obtain interesting, qualitative
information about the spectrum, which suggests unexpected
behavior near the QCD phase transition.

We begin with a normal, full QCD simulation at finite
temperature described by the partition function:
\begin{equation}
Z_{m,\beta} = \Pi_l\int d[U_l] e^{-\beta S_g}
\det(D+m)^{N_f/4}.
\end{equation}
\label{eq:part}
Here the integral is performed over all link matrices $U_l$ and
the gauge action $S_g$ is given by
\begin{equation}
S_g = -{1\over 3}\sum_{\cal P} {\rm re}\;{\rm tr}\;U_{\cal P}
\end{equation}
with $U_{\cal P}$ the ordered product of the four matrices
associated with the fundamental plaquette $\cal P$. We use the
staggered fermion Dirac operator $D$.

The determinant entering the partition function in
Eq.~\ref{eq:part} represents the effects of quark loops, for
$N_f$ flavors of quarks of mass $m$.  We will refer to the
corresponding lattice quark fields as $\chi(x)$.  In this paper
we introduce a second species of fermions, $\zeta(x)$, with mass
$m_\zeta$ which will appear in observables but does not enter the
quark determinant.  These $\zeta$ fermions may be thought of as
quenched or $N_f=0$ quarks.  We then compute the expectation
value of the quantity $\zbz$ in the thermal background specified
by the partition function of Eq.~\ref{eq:part}.  Thus,
\begin{equation}
\label{eq:zbz}
\langle\bar\zeta\zeta(x)\rangle =
\lan\; \langle x |{1 \over (D + m_\zeta)}|x\rangle\;\ranf
\end{equation}
where the average $\lan \; \ranf$ is calculated using the
distribution of gauge fields in the QCD path integral.
Although this quantity is no easier to compute than the normal
chiral condensate $\acbc$, the mass $m_\zeta$ can be varied
without changing the mass $m$ which enters the Boltzmann factor
and which must be fixed during a particular Monte Carlo run.
Typically we study 25 values of $m_\zeta$.

This quantity is of special interest because it is a simple
transform of the Dirac eigenvalue density.  If we evaluate the
propagator in Eq.~\ref{eq:zbz} by inserting a complete set of
eigenfunctions of $D$, and averaging over $x$, we find
\begin{equation}
\azbz(m_\zeta) = 2m_\zeta\int_0^\infty
{\rho(\lambda) \over (\lambda^2 + m_\zeta^2)} d\lambda.
\end{equation}
\label{eq:spectrum}
Here we have exploited the $\gamma^5$ symmetry which makes the
Dirac spectrum symmetric about zero and introduced
$\rho(\lambda)$, the Dirac eigenvalue density per unit volume,
averaged over the thermal distribution of Eq.~\ref{eq:part}.  The
function $\rho(\lambda)$ depends on $\beta$ and the dynamical
quark mass $m$.

Thus, $\azbz$ is a transform of the Dirac spectrum, strongly
weighted toward eigenvalues $\lambda \sim m_\zeta$.  Note the
$U(1)$ chiral symmetry for the quark field $\zeta$, present in
the staggered fermion formulation, implies that $\azbz$ should
vanish as $m_\zeta \rightarrow 0$.  Spontaneous breaking of this
symmetry results if $\rho(0)\ne 0$.  In which case the explicit
factor of $m_\zeta$ in the numerator of Eq.~\ref{eq:spectrum} is
compensated by the linear divergence of the $\lambda$ integral as
$m_\zeta \rightarrow 0$.  The result is the Banks-Casher
formula\cite{Banks}: %
\begin{equation}
\azbz(m_\zeta=0) = \pi\rho(0).
\end{equation}
Thus, we argue that the function $\azbz(m_\zeta)$  is easy to
compute, contains information about the spectrum of Dirac
eigenvalues and is closely connected to spontaneous symmetry
breaking.  In fact, as we discuss below, the infrared behavior of
$\azbz(m_\zeta\rightarrow 0)$ also determines the character of
anomalous chiral symmetry breaking, a quantity difficult to study
directly with lattice fermions.

  \pagenumbering{arabic}
  \addtocounter{page}{1}

\section{ $N_f=2$ CHIRAL TRANSITION}

\subsection{Critical Behavior}
\begin{figure}[htb]
\vskip-1in
\epsfxsize=75mm
\epsffile{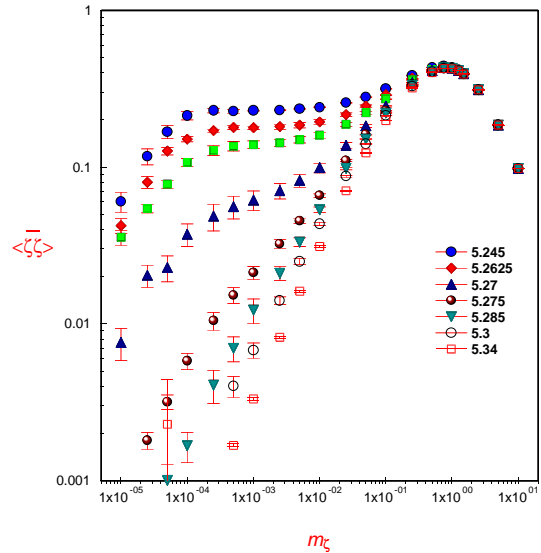}
\vskip-1cm
\caption{ The expectation value $\azbz$ plotted as a function of
the mass $m_\zeta$ for a series of values of $\beta$.  Shown are
two-flavor calculations with quark mass $ma=0.01$ on a $16^3
\times 4$ volume.
\label{fig:crit_bhv}
}
\end{figure}

Figure~\ref{fig:crit_bhv} shows the quantity $\azbz$ as a
function of $m_\zeta$ for a series of values of $\beta$ in the
critical region.  These results were obtained on a $16^3 \times
4$ lattice with two degenerate flavors of dynamical fermions of
mass $ma=0.01$.  Note, this is a log-log plot showing six orders
of magnitude variation in $m_\zeta$.  As can be seen, the
behavior is quite complex.

\begin{figure}[tb]
\vskip-0.5in
\epsfxsize=75mm
\epsffile{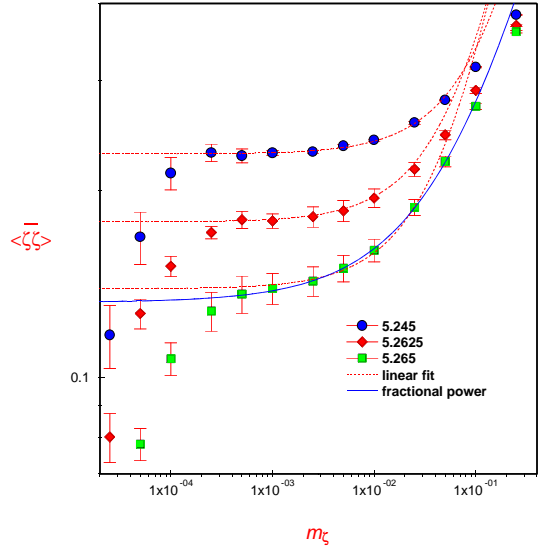}
\vskip-1cm
\caption{An expanded version of the previous $N_f=2$ plot showing
possible fits for $\beta \le \beta_c$.  These $\beta$ values all
show a non-zero intercept as $m_\zeta \rightarrow 0$, when
extrapolated from the region $m_\zeta \ge 5\times 10^{-4}$, a
region apparently unaffected by finite volume distortions.  Note
how the constant plus fractional power fit agrees with the
simulation over nearly three orders of magnitude for
$\beta=5.265$.
\label{fig:nf2_lpfl}
}
\end{figure}

First consider $\beta$ in the low temperature, chirally
asymmetric region, $\beta \le \beta_c$. For $m_\zeta \ge
10^{-2}$, $\azbz$ shows a complex non-linear behavior much of
which is certainly due to finite lattice spacing effects.  (In
the continuum limit a simple linear behavior is expected for
large $m_\zeta$ with a quadratically divergent coefficient.)  For
$10^{-4} \le m_\zeta \le 10^{-2}$ a constant plus linear fit
works well, at least for those values of $\beta$ farthest from
$\beta_c$.  For $m_\zeta \le 10^{-4}$ finite volume effects set
in and $\azbz$ approaches zero linearly as expected.
It is interesting to note that a somewhat more elaborate fit,
writing $\azbz$ as the sum of a constant and fractional power,
$\azbz = c_0 + c_1\,m_\zeta^y$  works well over a significantly
larger mass range $10^{-4} \le m_\zeta \le 0.05$ using a single
value of $y \approx 0.6$.  Such a fit merges nicely with a
description which holds at $\beta=\beta_c \approx 5.275$, where
only a fractional power, close to 0.6, is seen.
Figure~\ref{fig:nf2_lpfl} shows our $\beta=5.245$, 5.2625 and
5.265 data together with these constant plus linear and constant
plus fractional power fits.  For the smallest value,
$\beta=5.245$, the simple constant plus linear fit works nicely.
However, for the more nearly critical $\beta=5.265$,
Figure~\ref{fig:nf2_lpfl} shows the constant plus fractional
power working visibly better.

\begin{figure}[htb]
\vskip-.75in
\epsfxsize=75mm
\epsffile{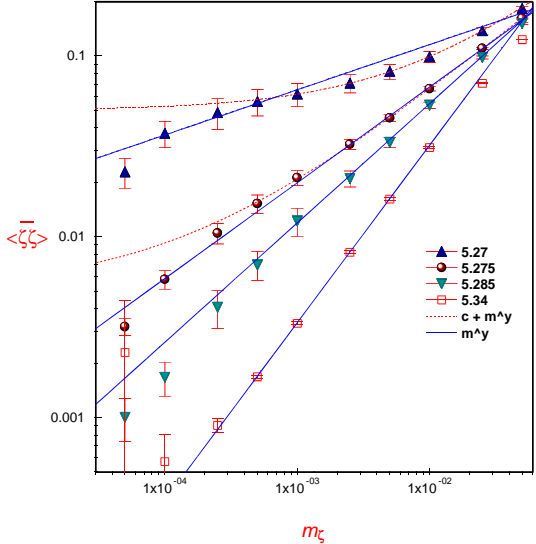}
\vskip-1cm
\vskip-0.25in
\caption{An expanded version of the previous $N_f=2$ plot showing
possible fits for $\beta \ge \beta_c$.  These $\beta$ values,
except perhaps for $\beta=5.27$, all show a zero intercept as
$m_\zeta \rightarrow 0$, when extrapolated from the region
$m_\zeta \ge 5\times 10^{-4}$.  Although a pure fractional power
does not fit very well at the smallest $\beta$ value shown, 5.27,
for larger $\beta$ this simple description works very well.
\label{fig:nf2_lpfh}
}
\vskip-.25in
\end{figure}

For $\beta \ge 5.275$, a fractional power, $\azbz=c\,m_\zeta^y$,
begins to fit well with the power growing with increasing $\beta$
so that by $\beta=5.34$ the power is $y=0.98$, very close to 1.0,
the expected linear dependence on $m_\zeta$ at high temperature.
Figure~\ref{fig:nf2_lpfh} shows these fits for $\beta=5.27$,
5.275, 5.285 and 5.34.  For the smallest $\beta$ value, 5.27, a
pure fractional power does not fit well---a constant term is also
needed while for the larger values of $\beta$ a pure fractional
power, with zero intercept, works quite well.  At these larger
values of $\beta$ the fits work at even lower masses suggesting
that the region of small eigenvalues, $\lambda \le \approx
10^{-4}$, which may be distorted by finite volume effects, is
less important.

If we attempt to study the critical behavior of $\acbc$ as a
function of the quark mass $m$, there are many predictions of
universality\cite{Detar}.  For example, at $T_c$, one expects
$\acbc = \kappa_1\,m^{1/\delta}$ for $m \rightarrow 0$. Of
course, this singular behavior in the mass can arise from a
combination of the dependence of $\acbc$ on the quark mass as it
enters both the propagator and fermion determinant.  This dual
role played by the quark mass can be seen explicitly by relating
$\acbc$ and $\azbz$:
\begin{equation}
\acbc|_{\beta,m} = \azbz(m_\zeta=m)|_{\beta,m}
\end{equation}
Here the mass in the quark propagator explicitly entering the
expectation value, $m_\zeta$, is often referred to as the valence
mass, $m_{\rm val}$, while the mass in the fermion determinant,
$m$, used when computing $\azbz$ is the sea quark mass, $m_{\rm
sea}$.

Thus, it is the dependence on $m_{\rm val}$ that we are
investigating when we examine $\azbz$ as a function of $m_\zeta$.
Only if the simulations do not depend in a critical way on the
quark mass in the determinant, should we attempt to compare our
exponent $y|_{\beta=\beta_c}$ with $1/\delta$.

\subsection{The High Temperature Phase}

The fact that $\azbz$ appears to vanish with vanishing $m_\zeta$
when $\beta \ge \beta_c$ is a bit surprising.  The conventional
picture assumes the presence of topological configurations that
break the $U_A(1)$ symmetry anomalously.  This requires a
significant density of small eigenvalues in the Dirac spectrum.
For example, a dilute instanton gas, expected by some to describe
the region above the critical point, would give a non-zero
density of eigenvalues, $\rho(0) \ne 0$. Recall, that it is only
for the physical quantity $\acbc$ that one expects the
restoration of chiral symmetry.  In the limit  $m \rightarrow 0$
not only is the valence mass in the propagator,  $m_{\rm val}$
vanishing, but also the mass in the determinant, $m_{\rm sea}$
is going to zero.  It is the added suppression of small
eigenvalues, coming from the zero quark mass limit of the fermion
determinant, that is presumably required for chiral symmetry
restoration above $T_c$.

Thus, a small sea quark mass should suppress the small
eigenvalues associated with anomalous breaking of $U_A(1)$
symmetry in the high temperature phase. Since in this calculation
we do not set  $m_{\rm sea}$ to zero, one might expect that
anomalous breaking of  $U_A(1)$ symmetry will require
$\azbz(m_\zeta\rightarrow 0) \neq 0$,  contrary to what we see.
Of course, this non-zero value of $\azbz$ will  be suppressed by
factors of $m_{\rm sea}a=0.01$. For that reason we will examine
larger dynamical quark masses and even a quenched calculation below.

\subsection{Anomalous $U_A(1)$ Symmetry Breaking}

Given the suppression of small eigenvalues that we see for
$\beta>\beta_c$ it is natural to ask if we can see any evidence
for anomalous symmetry breaking.  This is an important question
given the effects of such symmetry breaking on the order of the
two-flavor phase transition.

Consider a continuum, two-flavor theory with $SU(2) \times SU(2)$
flavor symmetry (in the limit of vanishing fermion masses) above
the phase transition.  The axial anomaly is expected to break the
anomalous $U_A(1)$, forcing the two-point functions constructed
from the operators $\bar\psi\gamma^5\tau^i\psi$ and
$\bar\psi\tau^i\psi$ to be unequal.  (Here $\psi$ is the
continuum fermion field.)  If these two-point functions are
evaluated at zero momentum and subtracted, their anomalous
$U_A(1)$ symmetry breaking difference (here called $\omega$) can
be written as follows:
\begin{equation}
\omega = 4m^2 \int_0^\infty {\rho(\lambda) d\lambda \over
          (\lambda^2 + m^2)^2}
\label{eq:omega}
\end{equation}
We can compare this equation with our results for the behavior of
$\rho(\lambda)$ as $\lambda \rightarrow 0$ at fixed quark mass.
In fact, the behavior of $\azbz$ as $m_\zeta \rightarrow 0$ for
$\beta$ just above $\beta_c$ suggests that $\rho(\lambda) \sim
c(m)\lambda^y$ for $0 < y <1$ which would suggest $\omega \sim
c(m)\,m^{(y-1)}$.  However, the fermion determinant may well
suppress the small eigenvalues, with $c(m)$ vanishing as $m$ goes
to zero.

\begin{figure}[htb]
\epsfxsize=75mm
\epsffile{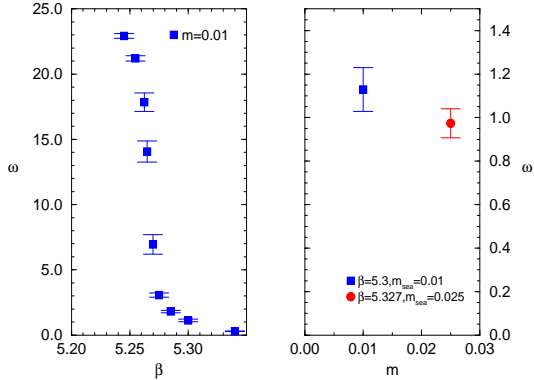}
\vskip-1cm
\caption{ The quantity $\omega$ which
directly measures the strength of anomalous symmetry breaking
plotted versus $\beta$.  On the left side results for $ma=0.01$
are shown while on the right, results for $ma=0.01$ and 0.025 are
compared at $\beta=5.3$ and 5.327 respectively.
\label{fig:omega}
}
\end{figure}

We can relate $\omega$ directly to our quantity $\azbz(m_\zeta)$:
\begin{equation}
\omega = {\acbc\over m} -
   {\partial\over\partial m_\zeta}\azbz|_{m_\zeta=m}
\end{equation}
and attempt to study how it varies as $m \rightarrow 0$.
Figure~\ref{fig:omega} shows the quantity $\omega$ for a series
of values of $\beta > \beta_c$ at $ma=0.01$ and at $\beta=5.327$
for $ma=0.025$.  Clearly $\omega$ is non-zero in this region and
does not appear to decrease as $ma$ decreases from 0.025 to 0.01
provided we adjust $\beta$ to remain a fixed displacement above
$\beta_c$.  Thus, this calculation appears to show the presence
of anomalous symmetry breaking for $N_f=2$, in the high
temperature phase, just above the transition.

\section{EFFECTS OF QUARK LOOPS}

\begin{figure}[htb]
\vskip-0.5in
\epsfxsize=75mm
\epsffile{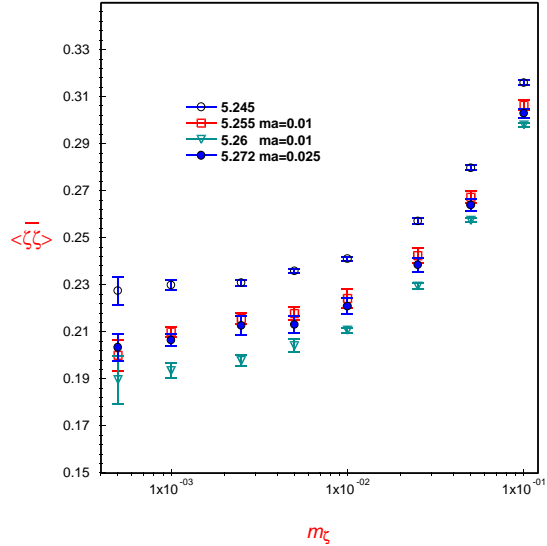}
\vskip-1cm
\caption{ The quantity $\azbz$ plotted as a function of $m_\zeta$
for $\beta=5.272$ and $ma=0.025$.  It is compared with $ma=0.01$
results for three values of $\beta$.  The close agreement with
the $ma=0.01$, $\beta=5.255$ curve is evident.
\label{fig:nf2_cmpl}
}
\end{figure}

\begin{figure}[htb]
\vskip-0.5in
\epsfxsize=75mm
\epsffile{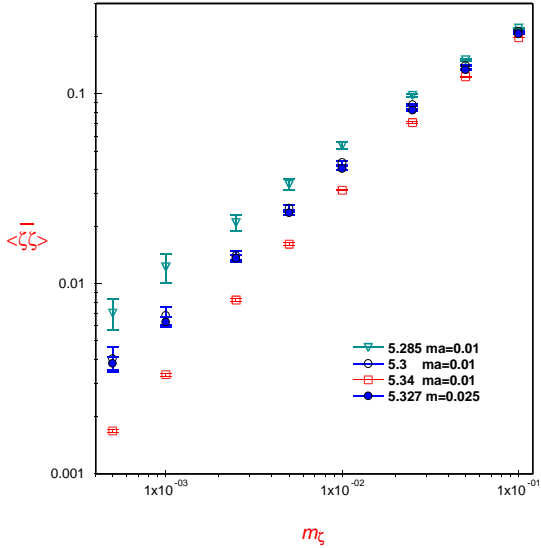}
\vskip-1cm
\caption{ The quantity$\azbz$ plotted versus $m_\zeta$ for
$\beta=5.327$ and $ma=0.025$, It is compared with $ma=0.01$
results for three values of $\beta$.  There is close agreement
with the $ma=0.01$, $\beta=5.3$ curve.
\label{fig:nf2_cmph}
}
\end{figure}

In order to address the effects of the fermion determinant, let
compare our $ma=0.01$ and 0.025 results in greater detail.  We
find that even when comparing the quantity $\azbz$ evaluated over
the entire range of $m_\zeta$, the $ma=0.01$ and $ma=0.025$
simulations can be made to agree within a few percent if we allow
for a quark-mass dependent shift in $\beta$.  In
Figure~\ref{fig:nf2_cmpl} we show $\azbz$ computed at
$\beta=5.272$, $ma=0.025$ compared to a group of $ma=0.01$
results.  The quite precise 5-6\% agreement with the
$\beta=5.255$, $ma=0.01$ curve over the entire $m_\zeta$ range is
striking.  Similarly, for higher temperatures,
Figure~\ref{fig:nf2_cmph} shows $\azbz$ computed at
$\beta=5.327$, $ma=0.025$ compared to  $ma=0.01$ results.  Again,
we see 5-6\% agreement with the $\beta=5.3$, $ma=0.01$ curve over
the entire $m_\zeta$ range.  Thus, in both the chirally symmetric
and asymmetric phases, the change in dynamical quark mass from
0.01 to 0.025 can be quite accurately compensated by a simple
shift in $\beta$.  No disparate effects on low or high
eigenvalues are seen.  The shift in $\beta$ of 0.027 needed in
the high temperature phase is precisely the shift in $\beta_c$
that we identified earlier when comparing $\beta_c$ found in our
0.01 and 0.025 simulations.

It is interesting to note that the 0.017 shift in $\beta$ found
in the low temperature regime is substantially smaller than the
0.027 shift needed at high temperature.  Thus, it is clearly
incorrect to describe the effects of quark loops as causing a
simple shift in $\beta$.  At the least, that shift is itself
$\beta$-dependent.

\section{QUENCHED CHIRAL TRANSITION}

\begin{figure}[htb]
\epsfxsize=80mm
\epsffile{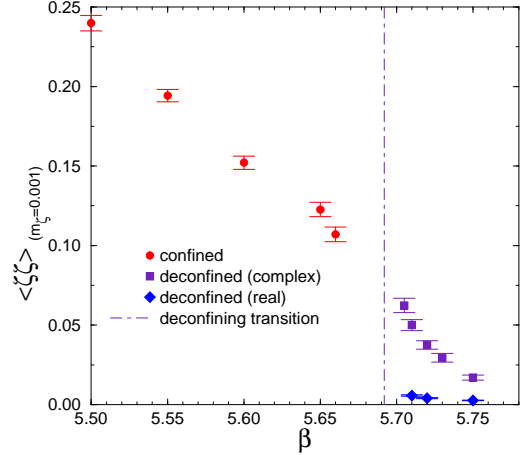}
\vskip-1cm
\caption{ The quantity $\azbz$ evaluated
at $m_\zeta=0.001$ is shown as a function of $\beta$ for pure
$SU(3)$ gauge theory.  The first-order, deconfining transition for
$N_t=4$ occurs at $\beta_c=5.6925$.  The multiple values for
$\azbz$ shown for $\beta \ge \beta_c$ correspond to the real and
complex phases with the smaller value of $\azbz$ occurring in the
real phase. \label{fig:qphase}
}
\end{figure}

As has been discussed above, we have not seen the effects of
small eigenvalues for $\beta > \beta_c$ in our $ma=0.01$, full
QCD simulations to the extent expected, for example, from a model
of dilute instantons.  In an attempt to enhance such possible
effects, we have repeated our calculation of $\azbz$ for the pure
gauge theory at and above the region of the deconfining phase
transition.  The results are summarized in
Figure~\ref{fig:qphase} where we plot the ``quenched'' chiral
condensate, $\azbz$ as a function of $\beta$ for a fixed mass
value, $m_\zeta=0.001$.

The results are quite surprising.  Because of the lack of $Z_3$
symmetry for the Dirac operator, $\azbz$ shows dramatically
different behavior in the three $Z_3$ phases for $\beta>\beta_c$.
The two phases where the Wilson line has a complex expectation
value (here called complex phases) show the same behavior for
$\azbz$ because they are related by complex conjugation.  This
complex phase shows spontaneous breaking of chiral symmetry, with
values of $\azbz$ that appear nearly continuous across the
deconfinement phase transition.  In contrast, the real phase has
a vanishing value of $\azbz$ in the limit $m_\zeta \rightarrow 0$
and looks very much like the chirally symmetric phase seen in our
full QCD simulations.

\subsection{Real phase}

\begin{figure}[htb]
\epsfxsize=80mm
\epsffile{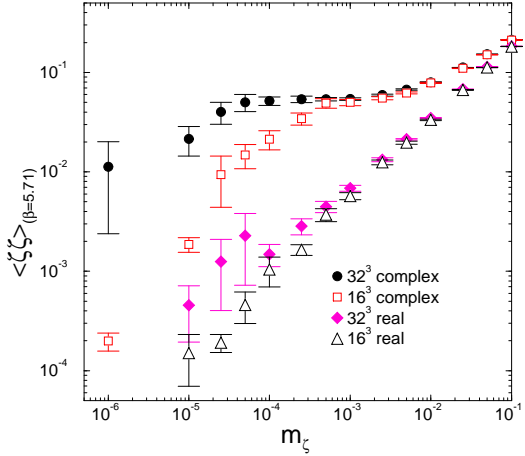}
\vskip-1cm
\caption{A comparison of $\azbz$ as a
function of $m_\zeta$ in the real and complex phases for
$\beta=5.71$ and $16^3$ and $32^3$ volumes.  The real phase shows
no perceptible volume dependence while the complex phase changes
significantly.
\label{fig:fig5_71}
}
\end{figure}

In Figure~\ref{fig:fig5_71} we show $\azbz$ in the real phase as
a function of quark mass, $m_\zeta$.  One sees a fractional power
dependence $\azbz = 0.99(3)\,m_\zeta^{0.73(1)}$, very much like
the behavior seen for $\beta>\beta_c$ in the full QCD
simulations.  As can be seen by comparing the results from the
$16^3$ and $32^3$ volumes, this behavior appears to characterize
the infinite volume limit.

\begin{figure}[htb]
\vskip-0.5in
\epsfxsize=75mm
\epsffile{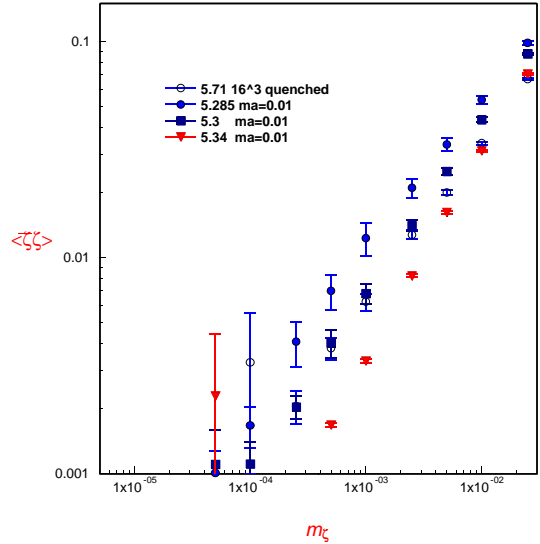}
\vskip-1cm
\caption{ A comparison of $\azbz$ plotted as a function of
$m_\zeta$ computed the real, quenched phase at $\beta=5.71$ (open
circles) with our $ma=0.01$, two-flavor results for a series of
values of $\beta$. Although the $\beta=5.34$, full QCD results
(solid inverted triangles) match most closely for larger quark
masses, as $m_\zeta$ decreases, the full QCD chiral condensate
falls more rapidly.
\label{fig:qch_cmph}
}
\end{figure}

Figure~\ref{fig:qch_cmph} compares the $m_\zeta$ dependence of
$\azbz$ in this real phase with that of our full $N_f=2$, ma=0.01
simulations.  The results are somewhat similar, showing perhaps a
15\% variation.  However, in contrast to the 0.01-0.025
comparison discussed earlier, there is a clear
systematic difference between the quenched and full QCD results.
As might be expected, the quenched results show an enhancement
for small eigenvalues not seen in the $ma=0.01$, two-flavor
calculation.  Never-the-less, we can describe the major effect of
the fermion determinant on QCD thermodynamics as selecting the
real phase (the phase with fewer small eigenvalues) as the
physical phase.

\subsection{Complex phase}

We see chiral symmetry breaking in the complex, deconfined phase
as might be expected from a dilute instanton gas model.  In
Figure~\ref{fig:fig5_71} we explicitly compare the dependence of
$\azbz$ on quark mass for $16^3$ and $32^3$ volumes.  As can be
seen, the presumed finite-volume ``knee'', where $\azbz$ abruptly
begins to decrease with decreasing quark mass dramatically shifts
to smaller quark mass on the larger volume.  The shift by roughly
the factor of 8 by which the volume has increased is exactly what
is expected for such spontaneous symmetry breaking.

\begin{figure}[htb]
\epsfxsize=75mm
\epsffile{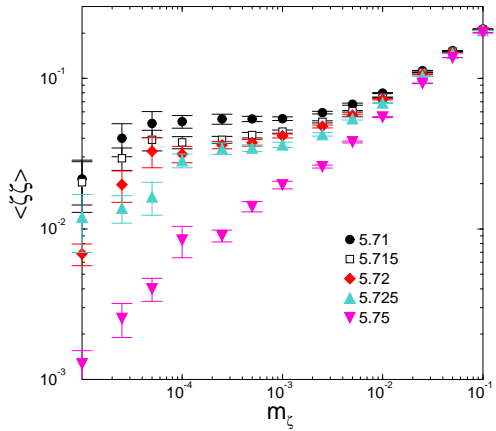}
\vskip-1cm
\caption{The quenched chiral condensate in the complex phase,
plotted versus $m_\zeta$ using only $32^3$ volumes for a series
of values of $\beta > \beta_c$.  The curves are well fit by a
constant plus a fractional power for $m_\zeta > 10^{- 4}$.
\label{fig:qch_cplx}
}
\end{figure}

\begin{figure}[htb]
\epsfxsize=80mm
\epsffile{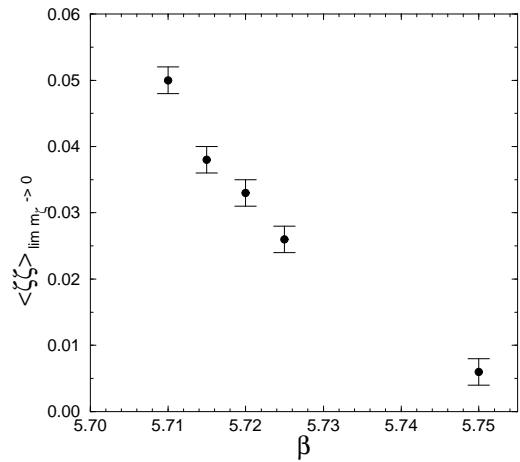}
\vskip-1cm
\caption{The constant term in our constant plus fractional power
fit, plotted against $\beta$.  We cannot distinguish with
certainty between the presence of a second, chiral symmetry
restoring transition for $\beta \approx 5.75$ and a rapid
decrease of $\azbz|{m_\zeta \rightarrow 0}$ as $\beta$ rises,
although the latter interpretation appears to be favored.
\label{fig:qch_extp}
}
\end{figure}

The variation of $\azbz$ with $m_\zeta$ is quite similar to that
seen for $\beta<\beta_c$ in our two-flavor simulations.  For
smaller $\beta$ a constant plus linear fit to $\azbz$ as a
function of $m_\zeta$ works quite well but as $\beta$ increases
and the constant intercept decreases, a constant plus fraction
power fit works significantly better.  This effect is best seen
on $32^3$ volumes and in Figure~\ref{fig:qch_cplx} we show
$\azbz$ versus $m_\zeta$ for a series of $\beta$ values.  This
similarity with our $\beta < \beta_c$, $N_f=2$ simulations
suggests the possibility of a second, chiral symmetry restoring
transition in $\azbz$ for this complex, deconfined phase.  In
Figure~\ref{fig:qch_extp} we plot the $m_\zeta \rightarrow 0$
values from the constant plus fractional power fits, versus
$\beta$.  Clearly we cannot rule out such a second phase
transition although a continuous, rapid decrease in
$\azbz|_{m_\zeta=0}$ as $\beta$ increases above the first-order
critical point appears the most natural interpretation of the
data.

\subsection{Inconsistent $N_f=1$ Phase Transition}

Combining our $N_f=2$ results with this $N_f=0$ calculation
suggests that the one-flavor phase transition may be quite
different from that expected on theoretical grounds, at least for
correspondingly coarse lattice spacings.  Since the density of
small eigenvalues found in the ``real'' deconfined phase for pure
QCD is insufficient to support a non-zero value of $\azbz$, the
$N_f=1$ high temperature phase, with increased suppression of
small eigenvalues coming from the fermion determinant, should
also show exact chiral symmetry with $\acbc=0$ in the limit $m
\rightarrow 0$.  Thus, for $N_f=1$, $\acbc_{m=0}$ is non-zero for
strong coupling but may well vanish for weak coupling which would
require that $N_f=1$ will show a phase transition.  This is in
conflict with the usual analysis which concludes that there is no
transition because of the anomalous breaking of the theory's
$U(1)$ chiral symmetry.

The failure to see the behavior expected for $N_f=1$ would be a
serious cause of concern either about our understanding of
continuum thermodynamics or about the coarse lattice spacings
used in present-day lattice calculations.

\section{CONCLUSION}

It should be noted that we have adopted a new approach to the
study of the axial anomaly using staggered fermions.  Direct
investigation of the axial anomaly is normally impeded by the
absence of a conserved, flavor-singlet axial current defined on
the lattice.  For example, a quantity like the $\eta'$ mass will
receive contributions from both chirally-asymmetric lattice
artifacts as well as the small Dirac eigenvalues of physical
interest.  By using operators with a different number of flavors
than appear in the Dirac determinant (operators which obey an
exact chiral symmetry on the lattice), we can study directly the
effects of the relevant Dirac spectrum---a quantity determined
solely by the number of flavors in the fermion determinant.  The
zero-momentum two-point function $\omega$ of Eq.~\ref{eq:omega}
examined for $N_f=2$ and the four-flavor chiral condensate
$\azbz$ discussed for $N_f=1$ are examples of this strategy.

\end{document}